\documentclass[11pt,twoside]{article}

\usepackage{asp2006}
\usepackage{epsf}
\usepackage{psfig}
\usepackage{lscape}

\begin{document}
\title{What Can We Learn from the Smallest AGN?}   %%% Fill in title
\author{Ari Laor}   %%% Fill in author names
\affil{Physics Department, Technion, Haifa 3200, Israel}    %%% Fill in author affiliations

\begin{abstract} %%% Abstract to run on from here.
Quite a few things. In particular, reverberation mapping of NGC~4395, the lowest
luminosity type 1 Active Galactic Nucleus (AGN, $L_{\rm bol}\sim 10^{40}$~erg/s) revealed a size of
only $\simeq 1$ light hour for the C~IV broad line region (BLR), which is by far the smallest BLR. 
This, together with a
similar determination of a size of $\sim 200$ light days in a luminous quasar 
(Kaspi et al. 2007), suggests that the $R_{\rm BLR}\propto L^{1/2}$ 
relation holds over a range of $10^7$ in $L$. This relation was suggested to
result from dust sublimation, which sets $R_{\rm BLR}$. This suggestion 
was beautifully confirmed recently by the dust reverberation results of
Suganuma et al. (2006). The $R_{\rm BLR}\propto L^{1/2}$ relation implies that the
broad lines width increases with decreasing luminosity according to $\Delta v\propto  L^{-1/4}$. 
But, there is an observational cutoff at $\Delta v\simeq 25,000$~km/s, and thus below
a certain threshold $L$ the BLR would not be detectable. Such objects constitute 
the so-called ``true type 2" AGN (e.g. most FR~I radio galaxies). The physical origin of the BLR gas is not 
established yet, but high quality Keck spectra of the H$\alpha$ profile in NGC~4395 
rule out a clumped distribution, and indicate that the gas resides in a smooth flow,
most likely in a thick rotationally supported configuration.
The H$\alpha$ line also reveals extended exponential wings, which are well modeled by 
electron scattering within the BLR emitting gas. Such wings can be used as a direct 
probe of the BLR temperature and optical depth.

\end{abstract}

\section{Introduction}   %%% Top level section head (remove "%" symbol)
Given an AGN spectrum, an experienced observer would be hard pressed
if asked to estimate the luminosity just based on the spectral emission features.
AGN spectra look remarkably similar over a range of as much as $\sim 10^7$ in luminosity.
The only significant trend is a decrease in equivalent width of most broad emission
lines with increasing luminosity, known as the Baldwin effect (Shields, these 
proceedings). But, even this trend is associated with a large scatter (e.g. fig.1 in 
Baskin \& Laor 2004). {\em Why are AGN emission line properties so weakly dependent on 
luminosity?  What is it telling us about the Broad Line Region size ($R_{\rm BLR}$)
and its luminosity dependence?} Determining the $R_{\rm BLR}(L)$ relation, and 
understanding its physical origin is important as the $R_{\rm BLR}(L)$ relation
is the cornerstone for the black hole mass ($M_{\rm BH}$) determination in AGN 
(e.g. Kaspi et al. 2005).
This relation also has important implications for the broad line region profile in low
luminosity AGN, and in particular for the absence of a BLR in very low luminosity AGN.
This review addresses the above questions together with new observations which shed some
light on their solution.

\section{The $R_{\rm BLR}(L)$ relation}
Preserving an overall similarity of the broad emission line spectrum
over a very wide range in luminosity requires a small range in the ionization parameter ($U$) 
and density ($n_e$) at the BLR, i.e. a small range in the photon density 
($n_{\gamma}=Un_e$) at the BLR.  Since $n_{\gamma}\propto L/R_{\rm BLR}^2$,
a small range in $n_{\gamma}$ indicates that $R_{\rm BLR}\propto L^{1/2}$. 
Direct measurements of $R_{\rm BLR}$ through reverberation mappings of the
H$\beta$ line indeed reveal $R_{\rm BLR}\simeq 0.1L_{46}^{0.56\pm 0.05}$~pc (Kaspi et al. 2005),
where $L_{46}$ is the bolometric luminosity in units of $10^{46}$~erg~$s^{-1}$.
Earlier studies (Kaspi et al. 2000) suggested a steeper relation, 
$R_{\rm BLR}\propto L^{0.69\pm0.05}$, but that was a result of using the optical 
(5100\AA) continuum luminosity density, which carries a smaller fraction of the bolometric
luminosity in more luminous objects. The slope of $0.56\pm0.05$ is obtained using 
the UV luminosity density, which lies closer to the peak of AGN spectral energy distribution.
{\em What is producing the} $R_{\rm BLR}\propto L^{1/2}$ {\em relation?} An attractive
possibility is that of Locally Optimally emitting Clouds (LOC, Baldwin et al. 1995), 
where gas at a wide range
of $U$ (and thus $n_e$) exists at all radii. The emissivity of each line is maximized
at a certain value of $U$ and $n_e$, i.e. at a certain $n_{\gamma}$. If the 
observed flux of each line is dominated by emission from gas with the peak emissivity, 
i.e  with a roughly constant $n_{\gamma}$, then the emitting gas will be located at 
$R\propto L^{1/2}$, as observed.
However, there is typically a broad range of $n_{\gamma}$ where the emissivity is close
to the peak value (e.g. Korista et al. 1997), which implies a correspondingly broad range 
in $R$. As a result, the radial dependence of the BLR gas covering factor becomes an important
effect, and the attractiveness of the LOC as a simple mechanism for a ``natural selection" 
of $R\propto L^{1/2}$ is lost. This problem is particularly significant for lines dominated 
by recombination, 
such as H$\beta$, as their emissivity is only weakly dependent on $n_{\gamma}$ (e.g. Korista et al.
1997), and it is mostly set by the fraction of absorbed photons, i.e. the local covering factor. 
Such lines will originate from any radius where there is gas 
exposed to the ionizing continuum, leaving the observed 
$R_{\rm BLR}\simeq 0.1L_{46}^{0.5}$~pc relation unexplained. An alternative mechanism,
where dust is responsible for the observed $R\propto L^{1/2}$ relation, was 
suggested by Laor \& Draine (1993), and Netzer \& Laor (1993), and is further discussed below.

Dust is clearly a major constituent of AGN, as indicated by its absorption (reddened AGN),
emission (the 3-100~$\mu m$ IR bump), and scattering (polarization) effects. Dust condenses out
of cooling gas (from stellar winds, or supernovae explosions) at a certain range of densities and temperatures, and it typically remains embedded in the gas
constituting about 1\% of the total mass. Dust generally survives as long as its temperature is below the sublimation temperature ($\sim$ 1400-1700K), which occurs at $R_{\rm sub}\simeq 0.2L_{46}^{0.5}$~pc 
for the most resilient grains (Laor \& Draine 1993). Dust sublimation
thus occurs interestingly close to $R_{\rm BLR}$, {\em is dust sublimation related physically to
the BLR size?}  

Dust embedded in photoionized gas suppresses the line emission
when $U>10^{-2}$, for the following reason. In the absence of dust, a photoionized slab of gas develops
a surface ionized H~II layer with a column density $\Sigma_{\rm ion}\simeq 10^{23}U$~atoms/cm$^2$. 
The optical depth of Galactic dust in the UV is $\tau\simeq \Sigma/10^{21}$, thus if the photoionized 
surface layer includes Galactic dust, then its optical depth in the ionized surface layer is
$\tau\simeq 100U$. The ionizing photons will hit a "dust wall" (i.e. reach $\tau>1$) once $U>10^{-2}$,
and they will not be able to penetrate further into the gas. Increasing $U$ will not increase significantly
the depth of the photoionzed layer, and the outgoing line flux will remain nearly constant. The excess
ionizing photons will be absorbed by dust and reprocessed into thermal dust emission. In addition, some of the outgoing line flux is also suppressed by dust absorption, in particular for resonance lines, as repeated scatterings increase their 
path length on the way out. Dust will also modify the line emission as some of the elements are heavily depleted into grains, and by affecting the thermal balance of the gas. Detailed photoionization calculations of dusty gas, including the above effects, indeed reveal a significant drop in the line emissivity once $U>10^{-2}$ (Netzer \& Laor 1993, Ferguson et al. 1997, Dopita et al. 2002). 

Observations indicate that at the narrow line region (NLR) $U\sim 10^{-3}$ to $10^{-4}$, while at the BLR 
$U\sim 0.1-1$. The NLR location is clearly distinct from the BLR, occurring in luminous AGN 
on scales of tens of pc or larger, vs. a tenth of a pc or smaller for the BLR. There is no significant line emission
coming from the region intermediate between the NLR and BLR, as indicated by the distinct profiles of the
broad and the narrow lines, and the lack of significant flux in lines with intermediate critical densities 
(e.g. the [O III]~$\lambda 4363$ line). {\em Is the intermediate region just devoid of gas?} clearly not, 
as it must contain a significant covering factor of dust close to the sublimation temperature, in order to
produce the observed $\sim 3-5~\mu m$ continuum. {\em Where is then the line emission from the gas which must be associated with the dust?}  A natural explanation is that $\log U>-2$ in the intermediate region, which suppresses the line emission. However, at $R<R_{\rm sub}$ the dust is destroyed, and cooling occurs only through line emission.
The BLR ``appears" once $R<R_{\rm sub}$. Dust sublimation provides both the correct luminosity
dependence and the absolute scale for $R_{\rm BLR}$. This mechanism involves no free parameters 
as dust sublimation is set by solid state physics. Also, this mechanism is essentially inevitable,
and the amount of suppression depends only on the assumption of Galactic dust/gas ratio. 

If correct, then the dust suppression mechanism implies that the $R_{\rm BLR}\propto L^{1/2}$ relation should 
extend to
all luminosities. To test that we performed reverberation mapping of NGC~4395, the lowest
luminosity type~1 AGN.

\section{Reverberation mapping of NGC~4395}
NGC~4395, discovered by Filippenko \& Sargent (1989), has 
$\nu l_{\nu}$(5100\AA)$\simeq 7\times 10^{39}$~erg/s, which 
makes it by far the lowest luminosity type 1 AGN (see Greene \& Ho 2004 for additional such objects).
It is $\sim 10^6$ less luminous than typical quasars, and $\sim 10^3$ less luminous than typical
Seyfert 1 galaxies, yet its UV and optical emission spectrum (Filippenko et al. 1993) 
is surprisingly similar
to that of normal type 1 AGN. The $R_{\rm BLR}(L)$ relation implies a remarkably small BLR,
of the order of one to a few light hours, and thus a direct measurement of $R_{\rm BLR}$ through reverberation mappings provides a significant leverage towards determining the slope of the 
$R_{\rm BLR}(L)$ relation. We were awarded 10 orbits with the {\em HST} STIS in April 2004 
to carry out reverberation mapping of the UV lines in NGC~4395 (PI-Laor). This program evolved
into a large campaign, including two space observatories and eight ground based observatories.
Specifically, ground based optical spectroscopy at the {\em KPNO} 4-m and {\em Lick} 3-m
telescopes, optical photometry at the {\em Wise Observatory} (1-m) at Mitzpe-Ramon, the {\em Nickel} (1-m) 
and {\em KAIT} (0.76-m) at the {\em Lick observatory}, IR photometry at {\em KPNO} (2.1-m), 
the {\em IRTF} (3-m) on Mauna Kea, and the {\em MAGNUM} (2-m) on Mt. Haleakala in Hawaii,
and X-ray spectroscopy for 60~ks with {\em Chandra}. The ground and space based observations were coordinated to overlap for $\sim 8$ hours per night on two nights at KPNO. Unfortunately, the program encountered unusually bad weather, and both Hawaii and {\em KPNO}
were clouded out during the campaign. The weather at {\em Lick} was clear, but the S/N of the optical spectroscopy there did not allow to detect significant variability of the Balmer lines. Even {\em HST} had problems, as only a single guide star was acquired for the first 5 orbits, allowing uncontrolled roll and
thus target drift, which compromised the required photometric precision. Fortunately, we were awarded 5 replacement orbits, carried out successfully in July 2004, just a month before STIS died. Significant variability was detected in the UV continuum and
in the C~IV line in each of the two 5 orbit visits, allowing two independent measurement of the lag. 
Both measurements were consistent with a $\sim 1$ hour delay, by far the shortest line delay ever measured. NGC~4395 thus has the most compact BLR, as expected for the lowest luminosity type 1 AGN.
The results of the UV, X-ray, and optical parts of the campaign are presented 
in Peterson et al. (2005), O'Neill et al. (2006), and Desroches et al. (2006). 

Earlier measurements of the C~IV delay were available for only four AGN (from earlier {\em IUE} and {\em HST} campaigns) 
confined to a range of $\log\nu l_{\nu}$(1350\AA)$=43.5-44$. These delays, together with the delays
measured for NGC~4395 in our campaign, where $\log\nu l_{\nu}$(1350\AA)$=39.6-40.1$, implied 
$R_{\rm BLR}\propto L^{0.61\pm 0.05}$ (Peterson et al. 2006). More recently, Kaspi et al. (2007)
measured the C~IV delay in a luminous $z=2.17$ quasar with  $\log\nu l_{\nu}$(1350\AA)=47, which together with the earlier measurements leads to $R_{\rm BLR}\propto L^{0.53\pm 0.04}$. The range in luminosity over which the C~IV delay is now measured increased
from a factor of 3 to a factor of $10^7$(!) within two years (compared to a range of $10^4$
available for H$\beta$, but with 34 rather than just 6 objects). Clearly, additional measurements
of the C~IV delay at both low and high luminosity are required, but the observations do appear to converge to a slope of 0.5, as expected if dust sublimation sets $R_{\rm BLR}$ in essentially 
all AGN.

\section{Dust Reverberation}

A direct way to test the dust-bounded-BLR scenario is by measuring the delay of the
near IR emission emitted by the hottest (and thus innermost) dust, in comparison to the delay of the broad
emission lines to optical/UV continuum variations in the same objects. Dust reverberation was first measured by
Clavel, Wamsteker \& Glass (1989) in Fairall 9, who found that the hottest dust lies ``at the
outskirts of the BLR" in this object. Few additional measurements of the size of the
IR emitting region were obtained following this study, but none included
a comparison with the size of the BLR in the same objects. Dramatic improvement followed from the Multicolour Active Galactic Nuclei Monitoring (MAGNUM) project, led by Y. Yoshii (IoA, Tokyo), Y. Kobayashi
(NAO, Tokyo), and B. Peterson (MSO, ANU), who built a 2-m telescope on Mt. Haleakala
dedicated to photometric monitoring of AGN in the U to K band range. A summary of the MAGNUM IR delay  results is given by Suganuma et al. (2006), which also includes a compilation of all earlier
IR delay results. The total sample includes 10 AGN covering a factor of $10^3$ in luminosity. The BLR delays of all objects were also compiled in this study. Two beautiful results came out of this
study, 1. $R_{\rm dust}\propto L^{0.5}$, as expected if dust always extends down to the sublimation
radius. 2. $R_{\rm dust}$ is just outside $R_{\rm BLR}$, as expected for the 
dust-bounded-BLR scenario. It therefore appears well established that dust sublimation controls the outer boundary of the BLR, and that dust suppresion of line emission is most likely responsible for the apparent gap between the broad and narrow line regions. 

{\em What then sets the inner boundary of the BLR?}  Since $U\propto n_e/R^2$, at $R\ll R_{\rm BLR}$
the gas will either 
go to the Compton temperature if $n_e$ remains fixed, leading to the loss of line cooling due to complete ionization, or, if $U$ remains constant, then $n_e$ has to increase inwards, leading to the loss of line cooling due to collisional deexcitations (e.g. Ferland \& Rees, 1988). Thus, {\em the location 
of the BLR and the NLR in AGN does not tell us where reprocessing gas is present, but rather where the gas cools mostly through line emission.}

Clearly, there is room for additional observational and theoretical work. In particular, extend the
range of luminosities where dust lags are determined (e.g. NGC~4395, Minezaki et al. 2006), and perform
a more careful theoretical study of dust sublimation as a function of the properties of the grains and the 
surrounding medium, as well as general studies of dust-gas interactions in the AGN environment.

\section{The BLR in typical low luminosity AGN}

NGC~4395 is unique in having an unusually low mass BH ($M_{\rm BH}\sim 10^5M_{\odot}$, 
Peterson et al. 2005) that results in rather normal looking broad line profiles despite the
low $M_{\rm BH}$. However, such low mass BHs appear to be uncommon (Greene, these proceedings),
and more typical values for $M_{\rm BH}$ are in the range of $10^7-10^9~M_{\odot}$. 
{\em How would the broad line profiles appear in more typical low luminosity AGN?}
Since the line width $\Delta v\simeq \sqrt{GM_{\rm BH}/R_{\rm BLR}}$, and $R_{\rm BLR}\propto L^{1/2}$,
we find that $\Delta v\propto M_{\rm BH}^{1/2}L^{-1/4}$. This relation implies that the broad lines
get broader with decreasing luminosity, a drop of $10^4$ in $L$ will result in 
a factor of 10 increase in $\Delta v$. {\em How broad can the lines get?} Observationally
lines broader than $\sim 25,000$~km/s are extremely rare. It is not clear whether this limit
is physical, i.e. if the BLR cannot exist above a certain $\Delta v_{\rm max}$,
or whether it is just a detection limit, as very broad lines may be hard to detect. One should note that 
in most cases where such broad lines are seen, they are very prominent 
(e.g. Gezari et al. 2007), suggesting
that the velocity limit may be a physical effect. Assuming $\Delta v_{\rm max}=25,000$~km/s implies 
that in objects with $L_{\rm bol}<L_{\rm min}=10^{41.8}(M_{\rm BH}/10^8M_{\odot})^2$,
or $L/L_{\rm Edd}<10^{-4.3}M_{\rm BH}/10^8M_{\odot}$, no
broad lines will be visible (Laor 2003). In such objects the dust sublimation radius shrinks to a size
where $\Delta v>\Delta v_{\rm max}$, leaving no room where the BLR can survive, or be detectable. 
Such objects will constitute the so-called ``true type 2" AGN, i.e. type 2 AGN where the BLR is truly lacking
rather than being obscured. Such objects are indicated in cases where there apparently exists an 
unobscured line of sight to the central compact continuum source (indicated by rapid optical/X-ray variability). 

The presence of ``true type 2" AGN was thought to be inconsistent with the inclination unification schemes, where all type~2 AGN are obscured type~1 AGN. But, the phenomenological presence of a 
$\Delta v_{\rm max}$ limit implies that
essentially all type~1 AGN turn into true type~2 AGN once $L_{\rm bol}<L_{\rm min}$.
Good candidates for true type~2 AGN are the FR~I radio galaxies,
which seldom show broad lines. The combination of high $M_{\rm BH}$ (related to their being radio loud), and low $L$ (related to their FR~I morphology), leads to very low $L/L_{\rm Edd}$, which puts most FR~Is
within the realm of true type 2 AGN. This is also indicated by the clear line of sight to a compact core 
in these objects. FR~II radio galaxies are significantly more luminous, and they indeed 
generally show broad lines. Interestingly, the handful of
FR~I galaxies with broad lines are all exceptionally luminous, indicating that the presence of a BLR is directly related to luminosity, rather than radio morphology (Laor 2003).

An alternative suggestion for a lack of broad lines in low luminosity AGN was put forward
by Nicastro (2000), in which the BLR is launched from a certain part in the accretion disk.
This part moves inwards with decreasing $L/L_{\rm Edd}$, and reaches the inner accretion
disk radius when $L/L_{\rm Edd}<(1-4)\times 10^{-3}$, at which point there will be no region
in the disk which can launch the BLR gas, and the BLR would no longer exist. This suggestion
implies that $\Delta v_{\rm max}$ can go up to $\sim 100,000$~km/s, and that the critical parameter
for the existence of a BLR is $L/L_{\rm Edd}$. Indeed, Nicastro et al. (2003) present observational 
evidence that only type~2 AGN exist at $L/L_{\rm Edd}\la 10^{-3}$. However, there are counter examples 
of type~1 AGN at $L/L_{\rm Edd}\sim 10^{-5}$, such as NGC~315, NGC~1052, NGC~1097, and NGC~3031 (Laor 2003; Lewis \& Eracleous 2006). It would thus be 
interesting to make a careful survey of broad lines in low luminosity AGN, which will be able
to detect very weak broad lines, in order to map out
the limiting $M_{\rm BH}$ and $L/L_{\rm Edd}$ below which a BLR is clearly not present. Such a study
can provide an interesting clue for the mechanisms which control the existence of a BLR.

\section{The NLR in low luminosity AGN}
{\em What happens to the NLR in low luminosity AGN?} 
In luminous AGN the NLR resides on scales of tens of pc, or larger, which is 
typically a scale where the dynamics is dominated by the bulge stars rather than by the BH, 
as indicated by the similarity of the stellar velocity dispersion and the [O~III]~$\lambda 5007$ 
line width (Nelson \& Whittle 1996). However, in low luminosity AGN the NLR becomes more compact,
and it may enter the region dominated by the BH gravity. The size of the [O~III]~$\lambda 5007$ emitting
region can be estimated as follows. The critical density for this
line is $n_{\rm crit}\sim 10^6$~cm$^{-3}$, and it is a dominant coolant for
$U\sim 10^{-3}$. This implies that the photon density at the most compact [O~III]~$\lambda 5007$ emitting region is $n_{\gamma}$(NLR)$\sim 10^{-6}n_{\gamma}$(BLR), and thus $R_{\rm NLR}\sim 10^3R_{\rm BLR}$.
The Keplerian velocity at the NLR due to the BH, neglecting the bulge, would then be a factor of $10^{1.5}$ smaller than in the BLR, giving a velocity dispersion of 
$\sigma_{\rm BH}=0.28 (M_{\rm BH}/M_{\odot})^{1/4}(L/L_{\rm Edd})^{-1/4}$~km/s (Laor 2003). According to
Tremaine et al. (2002), the bulge stellar velocity dispersion is  
$\sigma_*=1.9(M_{\rm BH}/M_{\odot})^{1/4}$~km/s,
and thus $\sigma_{\rm BH}/\sigma_*=0.15(L/L_{\rm Edd})^{-1/4}$. Therefore, the 
[O~III]~$\lambda 5007$ width can be set by the BH rather than the bulge once $L/L_{\rm Edd}<5\times 10^{-4}$. 
The NLR lines have a range of critical densities, and thus AGN with a low enough $L/L_{\rm Edd}$
can show a correlation of the line width with critical density, as beautifully demonstrated 
by Filippenko \& Sargent (1988) for NGC~3031, a LINER with $L/L_{\rm Edd}=5\times 10^{-5}$.
This study also reveals a transition from the expected $\sigma\propto n_{\rm crit}^{1/4}$ for 
lines with $n_{\rm crit}>10^5$~cm$^{-3}$, to $\sigma\sim$~constant for $n_{\rm crit}<10^5$~cm$^{-3}$,
corresponding to the transition from the BH dominated Keplerian velocity field, to the bulge 
dominated constant
velocity field. The Filippenko \& Sargent study inspired searches for a correlation of
$n_{\rm crit}$ and $\sigma$ in many luminous Seyferts (having $L/L_{\rm Edd}\sim 0.01-1$), 
but without any comparable success. The seminal discovery of the BH/bulge mass relation 10 years 
ago (Magorrian et al. 1998) now allows us to
understand that the NLR in all those objects lies entirely within the bulge potential,
explaining the lack of correlation of $n_{\rm crit}$ and $\sigma$ in those studies. 

The [O III]~$\lambda 5007$ line width is sometimes used as a proxy for $\sigma_*$ (e.g. Shields
et al. 2003), which together with the Tremaine relation provides an estimate of $M_{\rm BH}$.
The above implies that in objects with $L/L_{\rm Edd}\la 10^{-3}$ the [O III]~$\lambda 5007$ 
line may no longer be sampling the bulge potential, and lower ionization and lower $n_{\rm crit}$ lines
should be used instead (e.g. Greene \& Ho 2005). However, if the BH sphere of dominance can be spatially resolved
in low $L/L_{\rm Edd}$ objects, then the [O~III]~$\lambda 5007$ line can be used to trace the Keplerian
$v(R)$ from that region, and thus measure $M_{\rm BH}$ directly, as first achieved by Harms et al. (1994).

\section{What is the BLR made of?}
The physical thickness of the photoionized H~II surface layer is $d_{\rm ion}=10^{23}U/n_e$~cm.
Typical BLR values, $U=0.1$, $n_e=10^{10}$~cm$^{-3}$, give $d_{\rm ion}\sim 10^{12}$~cm, which is
much smaller than the measured range of sizes of the BLR, $10^{14}\le R_{\rm BLR}\le 3\times 10^{17}$~cm 
(Peterson et al. 2005; Kaspi et al. 2007). The small filling factor of the emission line 
gas in the BLR admits two possible configurations. The first is in the form of a clumped flow 
made of gas ``clouds", each one with a size $r_c>d_{\rm ion}$. An attractive source for
such clouds is the ``bloated stars" scenario, i.e. giants which are inflated by the incident
AGN continuum. This scenario provides a natural solution for the origin and possibly also survival
of the clouds. The required size of such stars is $r_*\simeq 10^{14}$~cm (e.g. Alexander \& Netzer 1994).
The second possible configuration is of a smooth gas flow
with a potentially large filling factor, such as a thick/warped disk or molecular clouds,
where the emission lines originate in a thin ionized surface layer. The two 
different configurations imply
different line profile characteristics. In the "clouds" configuration the observed broad lines
are the sum emission of all the clouds, where each cloud produces a narrow line which is only
thermally broadened. Statistical fluctuations in the number of clouds per velocity bin, $dn_c/dv$, 
will produce profile irregularities at a level of $1/\sqrt{dn_c/dv}$. In contrast, 
the "smooth flow" configuration leads to smooth line profiles. 

{\em How can we constrain the size of the BLR clouds?} High S/N and high spectral resolution 
observations of the broad line profiles can be used to measure/set limits on the profile irregularities. 
An upper limit on the profile fluctuations can then be converted into a lower limit on the required
total number of clouds in the BLR, $n_c$. This in turn can be used to place an upper limit
on the size of each cloud, $r_c$, through the following argument. The covering factor of the
BLR clouds is $C_{\rm BLR}=n_c\pi r_c^2/4\pi R_{\rm BLR}^2$. The strength of the broad lines
relative to the ionizing continuum implies $C_{\rm BLR}\simeq 0.1$. The two expresions for
$C_{\rm BLR}$ give 
$r_c\simeq R_{\rm BLR}/n_c^{1/2}$. Thus, a tight upper limit on $r_c$ can be obtained from
a high S/N spectrum of a low luminosity AGN. 
Such a limit was recently obtained through high quality Keck-II spectra of the H$\alpha$ 
profile in the lowest luminosity type~1 AGN, NGC~4395, which imply $n_c>10^4$ (Laor et al. 2006). 
Since $R_{\rm BLR}=10^{14}$~cm (Peterson et al. 2005), this implies $r_c<10^{12}$~cm.
This limit clearly excludes the ``bloated stars" scenario, which can in fact be excluded here
without resort to the fluctuation analysis, since $r_*\simeq R_{\rm BLR}$ and the bloated stars
are just too large to fit in the compact BLR of NGC~4395. Since $r_c\la d_{\rm ion}$,  
the BLR gas must effectively form a smooth flow. {\em What is the configuration of this flow?}
The BLR dynamics is dominated by gravity, and pure radial motions are excluded by reverberation
mapping (which show a symmetric line response). This leaves a thick 
(to get $C_{\rm BLR}\simeq 0.1$) rotating disk as a likely smooth flow configuration. 

{\em Why are the broad line profiles not always double peaked, as expected from gas in a rotating 
disk?}. Double peaked profiles are commonly seen in AGN with very broad lines (e.g. Gezari et al. 2007), but
most AGN show single peaked lines. There are, however, hints that some of the single peaked lines are actually strongly blended double peaks. This is indicated by distinct variability characteristics
of the blue and red wings in some objects (e.g. Wanders \& Peterson 1996), and a rotation of the
polarization angle from the blue to the red line wings (Smith et al. 2005). Apparently, there is a
relatively large velocity broadening mechanism in the BLR, which makes the double peaked structure
of a thin Keplerian disk blend into a single peak, but the individual peaks can still be discerned
through their distinct variability and polarization properties. When the Keplerian velocity is large enough the two peaks generally separate out. The velocity broadening should be in an ordered flow field
(to avoid shock heating of the gas through collisions), possibly in the form of a vertical wind from
a disk, which will also generate
the required $C_{\rm BLR}$. Reverberation mapping (i.e. maps of delay vs. velocity) would be 
valuable for a better understanding of the structure and dynamics of the BLR gas flow.

\section{Evidence for electron scattering in the BLR}
The very high quality Keck-II ESI spectrum of NGC~4395 (A. Barth, private communication), used
for the profile smoothness analysis, yielded another interesting result. The S/N in this spectrum
ranges from 50 per 0.259~\AA\ pixel at the continuum near H$\alpha$, to 400 at the H$\alpha$
line peak. This remarkably high S/N spectrum provides an H$\alpha$ profile of unprecedented accuracy, where
the shape of the far wings can be determined to $<10$\% error down to a flux density
of $10^{-3}$ of the peak flux density. At a level below 1\% of the peak flux density, which occurs at $v>1000$~km/s from the line center, the profile shows pure and symmetric exponential wings, i.e. 
$f_E\propto e^{-v/\sigma}$, with $\sigma=500$~km/s. Gaussian, Lorentzian, and Voigt profiles
are discussed extensively in various contexts, but no discussion of exponential profiles was
found in a search of the Astrophysics and Physics literature. A likely culprit appeared to be
optically thin electron scattering, based on the following simplified argument. Let $v_e$ be 
the scattering electron
velocity, then the typical energy gain of a scattered photon is $e_1/e_0\simeq 1+v_e/c$. After $n$
such scattering the photon energy will be $e_n/e_0\simeq (1+v_e/c)^n$, or $e_n/e_0\simeq 1+nv_e/c$
for $v_e/c\ll 1$. Changing from energy scale to velocity scale, defined through 
$v_n\equiv (e_n/e_0-1)c$, gives $v_n=nv_e$, i.e. each scattering increments the photon energy in
velocity space by $v_e$. For optically thin scattering, the fraction of photons scattered $n$
times is $\tau_e^n$, where $\tau_e$ is the optical depth. Thus, the photon flux per velocity bin
$f(v_n)=f(0)\tau_e^n$, or $f(v_n)=f(0)\tau_e^{v_n/v_e}$, which can be recast as 
$f(v_n)=f(0)e^{\ln \tau_e v_n/v_e}$, i.e. an exponential profile with $\sigma=-v_e/\ln \tau_e$.
Simply put, each scattering decreases the flux by a constant multiplicative factor, and increases
the energy by a constant linear factor, leading to an exponential profile. For Compton scattering, 
i.e. when $v_e/c\la 1$, the energy gain and the flux decrease per scattering are both
constant multiplicative factors,
and the above derivation leads to the well known power-law energy distribution (Rybicki \& Lightman 
1979, eq. 7.45). A more careful analytic derivation and numerical calculations provide the
following fitting function $\sigma=1.1v_e(-\ln \tau_e)^{-0.45}$ (Laor 2006). Thus, the slope of 
the exponential wings provides a constraint on $\tau_e$ and $v_e$, where $v_e$ is set by
$T_e$, the temperature of the scattering electrons. The optical depth $\tau_e$ is set
independently by the fraction of line flux in the scattering wings, and thus one obtains 
a unique solution for $T_e$ using the measured $\tau_e$ 
and $\sigma$. An acceptable fit to the H$\alpha$ profile in NGC~4395 
is obtained for $T_e=1.14\times 10^4$~K, and $\tau_e=0.34$. The formal fit errors on $T_e$ and
$\tau_e$ are tiny ($\sim 1$\%), but there is a larger error ($\sim 10$\%) related to
separating out the scattered line profile. {\em Where is the scattering gas located?}
Remarkably, $T_e$ and $\tau_e$ are both reasonably well reproduced for an optically thick photoionized 
slab with $U=0.3$, well within the plausible range of values for the BLR.
This indicates that the electron scattering takes place within the same photoionized layer where the
H$\alpha$ line is generated. These far exponential wings therefore provide us with a new way to
measure directly the temperature and optical depth of the BLR gas. The calculations of Laor (2006)
assume a simplified geometry and isothermal gas. Clearly, the H$\alpha$ emissivity changes with depth,
as does the electron temperature, so a more realistic radiative transfer calculation is required in
order to calculate the scattering line profile more accurately.

\section{Future prospects}

Observations of the emission properties of AGN over a very wide range in luminosity, in particular 
at the lowest luminosities, provide valuable hints for the physical origin and configuration
of the gas surrounding the massive black hole. Further progress can be made along the following lines.
\begin{enumerate}
\item Reverberation mappings of the BLR in other very low luminosity AGN, in addition to NGC~4395
(where the optical lines were not measured yet), to obtain statistically robust results.  
\item Extend dust reverberation to very low luminosity AGN, where the compact size may 
make it feasible to detect delays from cooler dust at wavelengths longer than the $K$ band.
\item Attempt reverberation mapping of the NLR, which may be detectable on timescales of 
months to a few years in low luminosity AGN ($L_{\rm bol}<10^{42}$~erg/s), assuming
there is significant variability on these timescales (e.g. NGC~1275).
\item Make a careful survey of broad lines in very low luminosity AGN, in order to map out
the limiting $M_{\rm BH}$ and $L/L_{\rm Edd}$ below which a BLR cannot exist. This will
provide important hints for the origin of the BLR gas.
\item Obtain ``true" Reverberation mappings, i.e. velocity resolved delay maps, of the BLR
for AGN at all luminosities, in particular those showing very broad/double peaked lines.
Such maps are important in order to constrain the structure and kinematics of the BLR.
\end{enumerate}
Complementary figures are not shown here due to lack of space. These can be obtained from the
presentation at the meeting, available at the meeting website, or at 
\verb physics.technion.ac.il/~laor .

\end{document}